\documentstyle[aps,multicol,prl]{revtex}
\begin{document}
\draft
\title{Exchange and Correlation Kernels at the Resonance Frequency -- Implications
for Excitation Energies in Density-Functional Theory}
\author{X. Gonze$^{\star,\dagger}$ and M. Scheffler$^{\star}$}
\address{$^{\star}$ Fritz-Haber-Institut der Max-Planck-Gesellshaft, Faradayweg 4-6,
D-14195 Berlin-Dahlem, Germany}        
\address{$^{\dagger}$ Unit\'e PCPM,
Universit\'e Catholique de Louvain,
B-1348 Louvain-la-Neuve, Belgium}
\maketitle
\begin{abstract}
Specific matrix elements of exchange and correlation kernels in time-dependent
density-functional theory are computed.
The knowledge of these matrix elements not only constraints approximate 
time-dependent functionals,
but also allows to link different practical approaches to excited states,
either based on density-functional theory, or on many-body perturbation theory,
despite the approximations that have been performed to derive them.

\end{abstract}
%



\begin{multicols}{2}[]

Density-functional theory (DFT), as proposed 
by Hohenberg, Kohn, and Sham~\cite{HKS}, has been highly successful in
the analysis of interacting-particle {\it ground states}.
Many efforts 
\cite{DeltaSCF,Theophilou79,Casida96Rev,Petersilka96a,Casida,Gora,Gorb,vanGisbergen98,Gorling,Filippi}
have also been devoted to DFT-based schemes
for {\it excited} states. 
The computation of excitation energies from 
DFT total-energy differences~\cite{DeltaSCF}, 
a procedure known as $\Delta$SCF, 
is formally restricted to transitions between the lowest states of different symmetries.
The ensemble-density-functional theory~\cite{Theophilou79}, a more general path
towards excitation energies, has not been used for actual 
calculations, to our knowledge.
Recently, the feasability of excitation-energy computations relying on 
two other, widely applicable, DFT-based formalisms has been demonstrated
for atoms and small molecules.
The first~\cite{Casida96Rev,Petersilka96a,Casida,Gora,Gorb,vanGisbergen98}
starts from the extension
of DFT to time-dependent (TD) phenomena~\cite{Runge84,vanLeeuwen98,approximateTDDFT} 
(TDDFT).
The second~\cite{Gorling,GorlingLevy,Filippi}, due to
G\"orling and Levy, builds
a perturbation theory (GLPT) in the difference between the many-body 
and the second-quantized Kohn-Sham (KS) Hamiltonians, 
where the parameter of the
perturbation is the coupling constant of the particle interaction,
in such a way that at each order, the exact density is recovered.

As an alternative to these DFT-based efforts, one may start from
many-body perturbation theory and perform partial resummations of diagrams
such as to build a screened interaction 
between dressed particles (quasi-particles)~\cite{HedinLundqvist}. 
At the lowest order in the screened interaction, one obtains 
Hedin's $GW$ approximation~\cite{Hedin65}
to quasi-particle energies (one particle is added or subtracted to the system), 
while the energy of excited states for which 
the number of particles is conserved can be deduced, in a subsequent step, 
from a Bethe-Salpeter (BS) equation describing
the interaction between quasi-particles (e.g. electron-hole pairs). 
The application of these techniques to 
real materials is very demanding~\cite{optical}.

In the present Letter we explore the relationships between 
excitation energies derived from time-dependent density-functional theory,
the G\"orling-Levy perturbation theory, 
and the screened-interaction many-body perturbation
theory. If excitation energies
were derived without approximations, in the three formalisms,
the results should be identical. Because of the specific
theoretical developments, the most obvious simplifications are different,
so that the practical schemes derived from these formalisms also differ.
We find that some approximations used for practical calculations 
leave a connection between the approaches,
at variance with the usual adiabatic local-density 
approximation (ALDA) in TDDFT~\cite{Casida96Rev,Petersilka96a,Casida}, 
that leads to a different physical picture.

The expressions from different approaches are linked thanks to a new technique
for computing selected elements of TD functional kernels at resonance, 
that is, at the frequencies corresponding to differences in KS
eigenvalues, for which the independent-particle susceptibility of the
KS system is resonant.
It is first applied to the {\it exact} exchange kernel, whose
matrix elements appear in a simplified TDDFT treatment of 
excitation energies based on a Laurent expansion, 
and are found {\it identical} to the first-order
corrections to KS eigenvalues differences, in the GLPT. 
The knowledge of these matrix elements imposes a new constraint on approximate functionals.
The same technique is then applied to an explicit XC functional 
that includes an approximate correlation contribution.
It is shown that the corresponding Laurent-expansion TDDFT excitation energies can be split 
according to powers of the screened interaction, as in the $GW$-BS approach,
and that they contain the $GW$ quasi-particle eigenvalue shifts.

For sake of simplicity, we will consider systems made 
of spinless particles (or equivalently, fully spin-polarized systems)
and focus on finite systems for which the energy levels
of interest are non-degenerate (KS wavefunctions
are real). The introduction of spin dependence and complex wavefunctions 
is technically straightforward, and does not modify the conclusions reached here.
We use Hartree atomic units throughout.
For integrals of the form 
$\int g_1({\bf r}_1) f({\bf r}_1,{\bf r}_2) g_2({\bf r}_2) d{\bf r}_1 d{\bf r}_2$,
in which $g_1$ and $g_2$ are {\it not} wavefunctions, but densities or products of
wavefunctions, we will use the notation $\{g_1|f|g_2\}$.

In TDDFT, the lowest-order Laurent expansion of the XC kernel
and KS susceptibility, 
in the difference between the actual excitation frequency, $\Omega$,
and the energy difference between corresponding KS
eigenenergies, $\omega^{\rm KS}_{\nu}=\epsilon^{\rm KS}_{j}-\epsilon^{\rm KS}_{k}$,
allows to deduce
approximate excitation energies as~\cite{Petersilka96a}:
\begin{eqnarray}
\Omega \approx \omega^{\rm KS}_{\nu}+\{ \Phi_{\nu} | V_{\rm C} | \Phi_{\nu} \}
    +\{ \Phi_{\nu} | f_{\rm xc}(\omega^{\rm KS}_{\nu}) | \Phi_{\nu} \} \quad.
\label{LaurentTDDFT}
\end{eqnarray} 
The double index $\nu$ = $(j,k)$ labels a single-particle transition from
the occupied level $k$ to the unoccupied level $j$, 
$\Phi_{\nu}({\bf r})$ is the product of 
KS wavefunctions $\phi_k({\bf r})\phi_j({\bf r})$, $V_{\rm C}$ is the Coulomb 
interaction kernel ($\frac{1}{|{\bf r}_1-{\bf r}_2|}$) 
and $f_{\rm xc}({\bf r}_1,{\bf r}_2;\omega)$ is the frequency-dependent XC kernel.

In Eq.(\ref{LaurentTDDFT}), 
the diagonal matrix elements of the Coulomb and XC
kernel for the transition between occupied and unoccupied states correct
the difference in KS eigenvalues.
The Coulomb kernel gives an exchange-like interaction between the KS electron
and hole
[$\phi_k({\bf r}_1)\phi_j({\bf r}_1) \frac{1}{|{\bf r}_1-{\bf r}_2|}
\phi_k({\bf r}_2)\phi_j({\bf r}_2)$]. 
The XC kernel is easy to evaluate in the ALDA 
(that ignore the functional dependence
of the XC potential $V_{\rm xc}$ 
on densities at different times as well as different locations):
It induces only a local, attractive, static electron-hole interaction
[$\phi_k({\bf r}_1)\phi_j({\bf r}_1) \delta ({\bf r}_1-{\bf r}_2)  
\frac{dV_{\rm xc}} {dn}({\bf r}_1)
\phi_k({\bf r}_2)\phi_j({\bf r}_2)$]. 
Then, as detailed by Cassida~\cite{Casida96Rev},
quasi-particle energies (i.e. excitation energies, to which
quasi-particle interactions make no contribution, like ionisation energies)
are identified with unmodified KS energies~\cite{NoteALDA}.

Unlike the ALDA, we evaluate exactly the exchange contribution. 
Since we will use later the same technique for the correlation contribution,
we treat them now on the same footing.
The Keldysh formalism, as applied recently
to the TDDFT case by van Leeuwen~\cite{vanLeeuwen98}
(we follow his notations), leads to the following explicit expression 
for the TD kernel, second derivative of the action $A_{\rm xc}$, 
in the temporal (pseudo-time) domain:


\begin{eqnarray}
f_{\rm xc}({\bf r}_1\tau_1;{\bf r}_2\tau_2)=&&
  \int \!\!\! \int_C dt_3 d{\bf r}_3 dt_4 d{\bf r}_4
\nonumber \\
 \frac{\delta u({\bf r}_3\tau_3)}{\delta n({\bf r}_1\tau_1)}
 && \frac{\delta u({\bf r}_4\tau_4)}{\delta n({\bf r}_2\tau_2)}
  \Big[
    \frac{\delta^2 A_{\rm xc}} {\delta u({\bf r}_3\tau_3) \delta u({\bf r}_4\tau_4)}
\nonumber \\
    -
    \int_C dt_5 d{\bf r}_5 && u_{\rm xc}({\bf r}_5\tau_5)
        \frac{\delta^2 n({\bf r}_5\tau_5)}
             {\delta u({\bf r}_3\tau_3) \delta u({\bf r}_4\tau_4)}
  \Big] \quad .
\end{eqnarray}
In this equation $u$ and $u_{\rm xc}$, respectively, denotes
the time-domain KS potential and XC potential.
 
This general expression can be specialized to the TD linear response
of an otherwise static system, in the frequency domain, and inserted in the
integral needed in Eq.(\ref{LaurentTDDFT}):
\begin{eqnarray}
\{ \Phi_{\nu} | f_{\rm xc}(\omega) | \Phi_{\nu} \}
  =
\{ \Phi_{\nu} &|& \chi_0^{-1}(-\omega)
 \nonumber\\
  \Big(
    \frac{\delta^2 A_{\rm xc}(\omega=0)}
         {\delta v_{\rm s}(-\omega) \delta v_{\rm s}(\omega)}
   &-&
    \int d{\bf r}_5 v_{\rm xc}({\bf r}_5)
        \frac{\delta^2 n({\bf r}_5;\omega=0)}
             {\delta v_{\rm s}(-\omega) \delta v_{\rm s}(\omega)}
  \Big) 
\nonumber \\
 && \chi_0^{-1}(\omega) | \Phi_{\nu} \} \quad .
\label{fxc_freq}
\end{eqnarray}
Note the appearance of the (static) XC potential $v_{\rm xc}({\bf r})$.
$\delta v_{\rm s}({\bf r};\omega)$ is a change
of KS potential in the frequency domain, $\chi_0({\bf r}_1,{\bf r}_2; \omega)$ is 
the independent-particle susceptibility of the KS system, and
$\chi_0^{-1}$ is the effective inverse of $\chi_0$, defined
on a restricted space excluding constant functions~\cite{Gorb}. 
The spatial dependence of potentials and susceptibilities
is implicit in Eq.(\ref{fxc_freq}), as well as
most integrals over spatial arguments.
$A_{\rm xc}(\omega=0)$ and $n({\bf r};\omega=0)$
are the time-independent part of the action integrand and the density, respectively
~\cite{NoteRectification}.

When evaluated at $\omega=\omega^{\rm KS}_{\nu}$,
Eq.(\ref{fxc_freq}) gives the XC kernel contribution
to $\Omega$ in Eq.(\ref{LaurentTDDFT}).
This requires the inversion of $\chi_0$ around the frequency $\omega^{\rm KS}_{\nu}$,
and its subsequent application to $\Phi_{\nu}({\bf r})$.
The explicit expression of $\chi_0$ is well-known
(see Eq.(14) of Ref.~\cite{Gorb}),
and exhibits a resonance at that frequency. The corresponding pole, can be singled out
before performing the inversion:
$
\chi_0({\bf r}_1,{\bf r}_2;\omega)=
  \frac{\Phi_{\nu}({\bf r}_1) \Phi_{\nu}({\bf r}_2)}
       {\omega-\omega^{\rm KS}_{\nu}}
  +
  \chi_{\nu}({\bf r}_1,{\bf r}_2;\omega) ,
\label{pole}
$
where the non-resonant contribution $\chi_{\nu}$ is implicitly defined by this equation.
The Sherman-Morrison formula~\cite{NumRecip} allows the analytical inversion of such
a decomposition. 
The application of $\chi_0^{-1}$ to $\Phi_{\nu}$, denoted $v_{\nu}$, gives:
\begin{eqnarray}
v_{\nu}(\omega)&=&  \chi_0^{-1}(\omega) | \Phi_{\nu} \}
\nonumber\\
&=&
\left( \omega-\omega^{\rm KS}_{\nu} \right)
 \frac {
 \chi_{\nu}^{-1}(\omega) | \Phi_{\nu} \}
       }
       { (\omega-\omega^{\rm KS}_{\nu}) -
         \{ \Phi_{\nu} | \chi_{\nu}^{-1}(\omega) | \Phi_{\nu} \}
       } \quad .
\label{anti-resonance}
\end{eqnarray}
A similar expression involving also  $\omega-\omega^{\rm KS}_{\nu}$ can be
derived in the case of $v_{\nu}(-\omega)$.
The presence of the factor $(\omega-\omega^{\rm KS}_{\nu})$ 
emphasizes the anti-resonant behaviour of $\chi_0^{-1}$.
This characteristics of Eq.(\ref{anti-resonance}) allows to extract the
leading order of the frequency expansion of Eq.(\ref{fxc_freq})
around $\omega^{\rm KS}_{\nu}$, for XC actions whose explicit expression
is known as a function of the KS potential through KS wavefunctions.
Indeed, after technically straightforward calculations within perturbation
theory (up to second-order wavefunctions), one finds
 
\begin{eqnarray}
\{ v_{\nu}(-\omega) &|&
     \frac{ \delta^2 n({\bf r};\omega=0) }
          { \delta v_{\rm s}(-\omega) \delta v_{\rm s}(\omega) }
   | v_{\nu}(\omega) \}
=
\nonumber\\
 && \phi_j({\bf r})\phi_j({\bf r}) - \phi_k({\bf r})\phi_k({\bf r})
 + {\cal O}(\omega-\omega^{\rm KS}_{\nu}) \quad ,
\label{density}
\end{eqnarray}
and, using the explicit exchange action of Ref.~\cite{vanLeeuwen98},
\begin{eqnarray}
\{ v_{\nu}(-\omega) &|&
    \frac{ \delta^2 A_{\rm x}(\omega=0) }
          { \delta v_{\rm s}(-\omega) \delta v_{\rm s}(\omega) }
  |  v_{\nu}(\omega) \}
 = 
\nonumber\\
  &&  \langle \phi_{j} | \hat{v}^{\rm HF}_{\rm x} | \phi_{j} \rangle
    - \langle \phi_{k} | \hat{v}^{\rm HF}_{\rm x} | \phi_{k} \rangle
   -
   \{ n_{k} | V_{\rm C} | n_{j} \}
\nonumber\\
  && + {\cal O}(\omega-\omega^{\rm KS}_{\nu}) \quad ,
\label{exchange}
\end{eqnarray}
where $\hat{v}^{\rm HF}_{\rm x}$ is the Fock non-local operator evaluated with KS
wavefunctions
and $n_{k}({\bf r})$ is the square of KS wavefunction $\phi_k({\bf r})$.
Combining Eqs.(\ref{fxc_freq}),(\ref{density}) and (\ref{exchange}), we derive the following
matrix element of the exchange kernel at resonance (in the limit of
$\omega \rightarrow \omega^{\rm KS}_{\nu}$), valid for 
all pairs $(j,k)$ of unoccupied and occupied levels associated
with a composite index $\nu$:

\begin{eqnarray}
\{ \Phi_{\nu} | f_{\rm x}(\omega^{\rm KS}_{\nu}) | \Phi_{\nu} \} 
&=&
       \langle \phi_{j} | \hat{v}^{\rm HF}_{\rm x} - v_{\rm x}  | \phi_{j} \rangle
     - \langle \phi_{k} | \hat{v}^{\rm HF}_{\rm x} - v_{\rm x}  | \phi_{k} \rangle
\nonumber\\
    &-& \{ n_{k} | V_{\rm C} | n_{j} \} \quad .
\label{exactfx}
\end{eqnarray}
 
{\it This explicit expression for selected matrix elements of the TD exchange kernel
provides, as such, a new constraint on approximate TD exchange
functionals}. 
Note the drastic simplification from the general matrix
elements of the TD exchange kernel derived by G\"orling~\cite{Gorb}.
The combination of Eqs.(\ref{LaurentTDDFT}) and (\ref{exactfx}), 
neglecting the correlation contribution, gives 
\begin{eqnarray}
\Omega \approx \omega^{\rm KS}_{\nu} &+&
       \langle \phi_{j} | \hat{v}^{\rm HF}_{\rm x} - v_{\rm x}  | \phi_{j} \rangle
     - \langle \phi_{k} | \hat{v}^{\rm HF}_{\rm x} - v_{\rm x}  | \phi_{k} \rangle
\nonumber\\
    &+&
    \{ \Phi_{\nu} | V_{\rm C} | \Phi_{\nu} \}
   -     \{ n_{k} | V_{\rm C} | n_{j} \} \quad .
\label{GLPT}
\end{eqnarray}
This expression is {\it identical} to the one obtained in the first-order
GLPT (see Eq.(A3) of Ref.~\cite{Filippi}, and also Ref.~\cite{Gorling}): 
{\it Excitation energies from the TDDFT with exact treatment of the exchange
kernel, in the Laurent
approximation, and from first-order GLPT are equal }~\cite{twoelectrons}. 
 
The KS eigenenergies $\epsilon^{\rm KS}_{j}$ and $\epsilon^{\rm KS}_{k}$
are expectation values of the KS Hamiltonian, that 
includes the exchange potential. Hence
the two expectation values of the difference between the Fock operator
and the exchange potential in Eq.(\ref{GLPT}) induce a shift of KS
energies $\epsilon^{\rm KS}_{j}$ and $\epsilon^{\rm KS}_{k}$, 
present in $\omega^{\rm KS}_{\nu}$,
in the direction of the Hartree-Fock quasi-particle energies.
So, {\it unlike in the ALDA approach, the quasi-particle energies 
from TDDFT with exact treatment of the exchange kernel 
are shifted from the KS eigenenergies.} 
The fourth term in Eq.(\ref{GLPT}) describes an unscreened Coulomb
electron-hole attraction, {\it to be contrasted} with the effective
interaction present in ALDA. However, the third term of Eq.(\ref{GLPT}) 
is {\it the same} exchange-like interaction as in the ALDA case, 
present in Eq.(\ref{LaurentTDDFT}).

The dynamical treatment of correlation is more difficult than that of exchange. 
We will not pursue the investigation of GLPT, that would involve
higher-order terms. However, we will establish
a link between TDDFT and
the screened-interaction many-body perturbation theory. At variance
with the decoupling between $GW$ and BS procedures, TDDFT 
computes the excitation energies in one operation. This
might be a crucial advantage for practical applications. Our aim will be
to show, at the adequate approximation levels, that TDDFT {\it contains} the
ingredients characteristics of the $GW$ approach, and to point out 
how it accounts for the additional BS terms.

The $GW$ quasi-particle shift, evaluated in a non-self-consistent, 
diagonal approximation, based on KS eigenenergy and eigenfunctions, leads to the
following excitation energies:
\begin{eqnarray}
\Omega \approx \omega^{\rm KS}_{\nu} 
  &+&    \langle \phi_{j} | \Sigma^{GW} 
                 (\epsilon^{\rm KS}_{j}) - v_{\rm xc}  | \phi_{j} \rangle
\nonumber\\
  &-&   \langle \phi_{k} | \Sigma^{GW} 
                 (\epsilon^{\rm KS}_{k}) - v_{\rm xc}  | \phi_{k} \rangle \quad .
\label{GW-BS}
\end{eqnarray}
 $\Sigma^{GW}$ is the self-energy operator in the $GW$ approximation:
\begin{eqnarray}
\Sigma^{GW} ({\bf r}_1,{\bf r}_2;\epsilon)= 
   \! \int \frac{d\epsilon'}{2 \pi} e^{-i\delta\epsilon'}
   iG ({\bf r}_1,{\bf r}_2;\epsilon-\epsilon')
   W ({\bf r}_1,{\bf r}_2;\epsilon') ,
\nonumber
\end{eqnarray}
where the Green's function $G$ is built from KS wavefunctions, and $W$
is the dynamically screened Coulomb interaction 
($W=\varepsilon_{\rm RPA}^{-1} V_{\rm C}$, 
with $\varepsilon_{\rm RPA}=1- V_{\rm C}\chi_0$ being 
the RPA dielectric screening).
The $\epsilon'$-integral 
can be further transformed to a contour integral in the complex plane.

In Eq.(\ref{GW-BS}), the difference between KS eigenenergies 
is shifted by expectation values
of the difference between the self-energy operator and the XC
potential. 
The neglected BS corrections would make quasiparticles interact
through an unscreened exchange-like term (compare with Eq.(\ref{LaurentTDDFT}))
as well as through a {\it screened} Coulomb interaction.
Excitonic effects would be described at that level. The screened
Coulomb interaction is
to be contrasted 
with the {\it effective}, {\it local}, electron-hole interaction 
of ALDA-TDDFT, and the {\it unscreened} Coulomb interaction of first-order GLPT. 

A formally exact correlation functional can be
obtained by the combination of the adiabatic connection method (based on a
coupling constant integral) and the fluctuation-dissipation theorem~\cite{Dobson}. 
In the context of the DFT computation of van der Waals energy, Dobson~\cite{Dobson}
proposed a simplified expression, 
\begin{eqnarray}
E_{\rm c}=\int_0^1 d\lambda
     \int_{0}^{\infty} \frac{du}{2\pi}
     {\rm Tr} \left[ V_{\rm C} ( \chi_0(iu) - \chi_{{\rm RPA},\lambda}(iu) ) 
              \right] \quad ,                            
\label{ec}
\end{eqnarray}
in which the true susceptibility
present in the exact $E_{\rm c}$ expression is replaced by the RPA one,
although the integral over the coupling constant $\lambda$ is retained.

The RPA susceptibility 
is a simple functional
of $\chi_0$ [~$\chi_{{\rm RPA},\lambda}=\chi_0 (1-\lambda V_{\rm C}\chi_0)^{-1}$ ]. 
This latter quantity is a functional of the KS potential, itself
a functional of the density. We can thus compute
the corresponding correlation potential and kernel. Thanks to an integration by parts,
we obtain that the correlation potential does {\it not} 
exhibit the integration over the coupling constant $\lambda$:
\begin{eqnarray}
v_{\rm c}({\bf r}_1) 
 = \frac{\delta E_{\rm c}} { \delta n({\bf r}_1)}
 &=& \int d{\bf r}_2
     \chi_0^{-1}({\bf r}_1,{\bf r}_2;\omega=0) 
     \int_{0}^{\infty} \frac{du}{2\pi}
\nonumber\\
   &&  \times {\rm Tr} \left[ (V_{\rm C}-W(iu))
                     \frac{\delta \chi_0(iu)} {\delta v_{\rm s}({\bf r}_2)}
              \right]
     \quad ,
\label{vc}
\end{eqnarray}
where the same screened interaction $W$ as in the $GW$ approximation appear.
For the computation of excited states, we generalize the correlation functional
Eq.(\ref{ec}) to a TD action, and, neglecting
a term in the second power of the screened interaction, derive 
\begin{eqnarray}
&&\frac{\delta^2 A_{\rm c}(\omega=0)}
     {\delta v_{\rm s}({\bf r}_1;-\omega) \delta v_{\rm s}({\bf r}_2;\omega)}          
  \approx
\nonumber\\
&&\int_{0}^{\infty} \frac{du}{2\pi}
     {\rm Tr} \left[ (V_{\rm C}-W(iu)) 
                     \frac{\delta^2 \chi_0(iu)} 
                     {\delta v_{\rm s}({\bf r}_1;-\omega) \delta v_{\rm s}({\bf r}_2;\omega)}
              \right] \quad .
\label{Kc}
\end{eqnarray}  
For comparison with the $GW$ approximation, it is crucial that 
the coupling constant is absent, as in Eq.(\ref{vc}).
The second derivative of $\chi_0$ with respect to TD
changes of potential can also be evaluated (perturbation theory
up to third-order wavefunctions), close to the resonance:
\begin{eqnarray}
\{ v_{\nu}(-\omega) &|&
    \frac{ \delta^2 \chi_0({\bf r}_1,{\bf r}_2; iu)  }
          { \delta v_{\rm s}(-\omega) \delta v_{\rm s}(\omega) }
  | v_{\nu}(-\omega) \}
=
\nonumber\\
 && \phi_j({\bf r}_1) \,
  2 \, {\cal R}{\rm e} \left[ G({\bf r}_1,{\bf r}_2; \epsilon^{\rm KS}_{j} + iu) \right] \,
 \phi_j({\bf r}_2) 
\nonumber\\
&-& 
\phi_k({\bf r}_1) \,
  2 \, {\cal R}{\rm e} \left[ G({\bf r}_1,{\bf r}_2; \epsilon^{\rm KS}_{k} + iu) \right] \,
 \phi_k({\bf r}_2)
\nonumber\\
&+& {\cal O}(\omega-\omega^{\rm KS}_{\nu}) \quad ,
\label{d2chi0}
\end{eqnarray} 
where the same Green's function $G$ as in the $GW$ approximation appears.
The combination of Eqs.(\ref{Kc}) and (\ref{d2chi0}), needed in Eq.(\ref{fxc_freq}), 
gives the difference between an unscreened $GV_{\rm C}$ term and a $GW$
term. We introduce this result in Eq.(\ref{LaurentTDDFT}), and find that
the main contribution of the first term ($GV_{\rm C}$) cancels the 
unscreened Fock operator that shifts the KS eigenvalues in Eq.(\ref{GLPT}).

We can now discuss the link between $GW$-BS excitation energies 
and TDDFT excitation energies obtained
from an exact exchange kernel and an approximate correlation kernel :  
$i)$ as expected, the crucial $GW$ terms are contained in Eq.(\ref{LaurentTDDFT});
$ii)$ the inclusion of correlation in TDDFT transforms the {\it unscreened}
single-orbital matrix elements of Eq.(\ref{GLPT}) to {\it dynamically screened}
matrix elements, as in Eq.(\ref{GW-BS});
$iii)$ the subtraction of the same matrix elements of $v_{\rm xc}$ appear
in both schemes, compare Eq.(\ref{fxc_freq}-\ref{density}) to Eq.(\ref{GW-BS}); 
$iv)$ however the contour integral deduced from Eq.(\ref{Kc})
is shifted from the contour integral
in the $GW$ approximation, and includes contributions from additional poles ;
$v)$ a term in the second power of the screened interaction has been neglected
from Eq.(\ref{ec}) to Eq.(\ref{Kc});
$vi)$ the RPA susceptibility was used in Eq.(\ref{ec}).
Points $iv)$-$vi)$ connects to BS as well as ${\cal O}(W^n)$ (with $n\ge 2$)
corrections to Eq.(\ref{GW-BS}). Understanding point $iv)$
also requires a careful analysis of our TD correlation action
(especially the contour integral
to be used with the fluctuation-dissipation theorem).
A more detailed discussion of these aspects will be left for a future publication.

In this paper, we have examined the time-dependent
kernels deduced from the exact exchange functional and an approximate 
correlation functional, at resonance. Links between
approximate expressions for excitation energies 
from time-dependent DFT, the G\"orling-Levy perturbation theory,
and Hedin's $GW$ approximation have been found:
The difference between Kohn-Sham eigenvalues are 
corrected by one-particle shifts, followed by 
(unscreened or screened) interaction effects. 
We also noted that in the adiabatic 
local density approximation to time-dependent DFT, no 
one-particle shift is present, and the link with the two other techniques
is lost. 

\section*{Acknowledgments}
We thank A. Schindlmayr for a careful reading of the manuscript.
X.G. acknowledges financial support from the A. von Humboldt foundation and
from the National Fund for Scientific Research, Belgium.

\end{multicols}


\begin{references}
\bibitem{NoteALDA}
In the complex version of Eq.(\ref{LaurentTDDFT}) -- needed for scattering states --,
if one of the wavefunctions is delocalized, within ALDA, both Coulomb and XC
contributions vanish.
\bibitem{NoteRectification}
Although time-independent, they are modified by a TD perturbation,     
in the same way as an oscillating current induces a constant potential drop
in a diode, a phenomenon known as rectification. 
Albeit not apparent in our notation, 
perturbations in the frequency domain are switched on adiabatically, as
in Ref.~\cite{Gorb}.
\bibitem{twoelectrons}
For the specific case of two-electron systems,
this equality had been noticed 
(p.9999 of Ref~\cite{Filippi} ; p.2565 of Ref.~\cite{vanGisbergen98}),
but it was not suspected to hold in general.
\end{references}
\end{document}